\documentstyle[rmaaconf]{article}

\input epsf

\begin{document}

\newcommand{\cidfig}[6]{
    \protect\centerline{
    \epsfxsize=#1\epsffile[#2 #3 #4 #5]{#6}
     }}
\newcommand{\etal}{et al.}
\newcommand{\ldo}[1]{\hbox{$\lambda$#1 \AA}}
\def\lsim{\mathrel{\lower2.5pt\vbox{\lineskip=0pt\baselineskip=0pt
           \hbox{$<$}\hbox{$\sim$}}}}
\def\gsim{\mathrel{\lower2.5pt\vbox{\lineskip=0pt\baselineskip=0pt
           \hbox{$>$}\hbox{$\sim$}}}}
\def\Hb{\mbox{H{\small $\beta$}}}
\def\ewHb{\hbox{$W_{\mbox{\scriptsize H{\tiny $\beta$}}}$}}
\def\ojo{\fbox{\bf !`$\odot$j$\odot$!}}   
\def\Zsun{\hbox{Z$_{\odot}$}}
\def\uniden{\hbox{cm$^{-3}$}}
\def\LBclu{\hbox{$L_B^{clu}$}}
\def\uniSNrate{\hbox{yr$^{-1}$}}
\def\SNrate{\hbox{$\nu_{SN}$}}
\def\LBstar{\hbox{$L_B^{\star}$}}
\newcommand{\nuLB}{$\SNrate / \LBstar$}
\def\eB{\hbox{$\epsilon_B$}}
\def\n7{\hbox{$n_{7}$}}
\def\LBsun{\hbox{$L_B^{\odot}$}}
\def\tsg{\hbox{$t_{sg}$}}
\def\e51{\hbox{$\epsilon_{51}$}}
\def\ni{\noindent}                                       
\def\Msun{\hbox{M$_{\odot}$}}

\title{AGN variability: from Seyfert nuclei to QSOs}

\author{Itziar Aretxaga\altaffilmark{1}}
\altaffiltext{1}{Royal Greenwich Obs. Madingley Rd. Cambridge, CB3 0EZ, UK}

\begin{resumen}
  Las propiedades del continuo \'optico de los N\'ucleos Gal\'acticos Activos
son revisadas, encontr\'andose que las variaciones 
producidas en objetos d\'ebilmente radioemisores son consistentes con 
aquellas esperadas en modelos Poissonianos, en los 
que la luminosidad es el producto de la superposici\'on 
de pulsos individuales. En este contexto se describen las predicciones del
modelo de Formaci\'on Estelar Violenta.
\end{resumen}

\begin{abstract}
The continuum variability of optically selected Active Galactic
Nuclei (AGN) is found to be consistent with that expected from a simple
Poissonian process, in which the total luminosity of an object is produced 
by the multiple superposition of identical pulses. The energies, 
time-scales and
rates of the pulses are found to be in the range of those expected 
from supernovae which generate fast evolving remnants in a nuclear 
starburst.
However, radio-loud AGN 
don't follow the predictions of that simple 
scenario.
\end{abstract}

\section{Introduction}

Variability is a common characteristic of AGN. Even among brands which
classically have been regarded as quiescent, LINERs and Seyfert~2 nuclei,
there are reported cases of variations in both lines and continuum ---  
see for example the development of broad lines and increase of continuum 
luminosity
in the LINER NGC~1097 (Storchi-Bergmann \etal\ 1993) or in the Seyfert~2 
nucleus Mrk~993 (Tran \etal\ 1992). Classically variable brands
are Seyfert~1 nuclei and QSOs, with simultaneous time-scales of variation 
which range from a few years to a few weeks (see the sketch in Smith \etal\
1991) and involve integrated $B$-band energies of up to several times 
$10^{50}$~erg. 
In some of these radio-quiet AGN intra-day variability has also 
been found (Dultzin-Haczyan \etal\ 1992, Gopal-Krishna \etal\ 1995), a mode 
which is common in blazar-type objects (Wagner et al. 1990).

Continuum variability is a popular tool to assess
models of AGN and, although detailed predictions are 
rarely available, there is considerable debate  
about trends in the variability relationships 
which can rule out or backup one model or another.

\section{QSO variability from ensemble light curves}

Much of our knowledge about QSO variability is based on studies of
large samples of sources monitored on photographic plates over a
period of one or two decades. Although each individual QSO is poorly
monitored in these samples (typically 10--20 epochs), the rationale 
is that the ensemble light curve of all QSOs will give general
information about the individuals.
In the last decade several studies have shown that
variability is anti-correlated with luminosity, in the sense that
luminous QSOs have smaller amplitude variations than low-luminosity ones 
(e.g.
Pica \& Smith 1983, Hook \etal\ 1994). The anti-correlation is
somewhat flatter than a ``$1/\sqrt{N}$'' law, a result which seems to
rule out simple Poissonian models, in which the variations are created
by a random superposition of identical events, or pulses.

\subsection{Danger in wavelength effects}

However, these studies disregard the fact that AGN variability is wavelength
dependent. It is known that in nearby AGN the
amplitude of variations increases towards shorter wavelengths. This
is illustrated in Kinney \etal\ (1991) where the continuum
variations across the UV spectrum are shown for all the AGN in the IUE 
database. A quantitative study of the Seyfert~1 nuclei and QSOs ($z<1.3$) in 
this sample states that in the
\ldo{1200--3200} range the rms of the luminosity changes by about 
($6.2\pm4.3$)\% every 1000~\AA\ 
(Paltani \& Courvoisier 1994). Indirect evidence that
wavelength effects are also present in high-redshift QSOs was derived from
parametric fits to single pass-band variability data 
(Cristiani \etal\ 1996, Cid Fernandes, Aretxaga \& Terlevich 1996,
hereafter CAT96); but the first
direct evidence for this effect has been presented by Cristiani
\etal\ in this conference, showing that the $R$-band amplitude
of variability in a
sample of 149 QSOs is smaller by a factor $1.13\pm0.05$ than in $B$-band.
 
The
variability measured at a fixed optical band tends to
overestimate the monochromatic rest frame optical variability, simply
because the objects at higher redshifts are observed at bluer emitted 
wavelengths.
Wavelength effects must be removed before analyzing the variability
dependence with luminosity and deriving implications for Poissonian models.
Indeed, parametric fits
to the variability--luminosity--redshift space
(CAT96) show that if QSOs in general follow a 
wavelength--variability relationship similar to that of nearby AGN 
($\sigma \propto \lambda^{-b}$, with $0.5 < b < 1.5$, where $\sigma$
is the rms of the luminosity), then the
variability-luminosity relationship $\sigma \propto L^{-a}$ 
can be bracketed to values $-0.8 < a < -0.3$ and, therefore, is consistent
with a simple Poissonian model ($a=-0.5$).
  
\subsection{Parameters for a simple Poissonian model}
 
 The variability generated by a Poissonian process
is characterized by the following parameters:\\
    \hspace*{1cm}    1. \ the percentage of a non-variable ``background''
                        component, if it exists ($f_{bck}$);\\
    \hspace*{1cm}    2. \ the time-scale of the events 
	                ($\tau$);\\
    \hspace*{1cm}    3. \ the energy of the events ($\epsilon$);\\
    \hspace*{1cm}    4. \ the rate of events ($\nu$).\\ 

An upper limit for the background fraction of light
can be directly estimated from the minima of the light curves of 
QSOs: $f_{bck} \lsim L_{min}/\overline{L}$. This
fraction is $f_{bck} < 0.7$ for the SGP sample,
283 QSOs at $0.43<z<4.01$ observed in $B$-band for 7 epochs in a time span 
of 16~yr  
(Hook \etal\
1994). 

  The time-scale of the pulses can be derived from the ensemble Structure
Function (SF) of QSOs. The SF is the curve of growth of variability with time,
formally defined  as $\mbox{SF}(\Delta t)=<m_i - m_j>$, where $i$ and $j$ are
 all the possible epochs which satisfy that $|t_i - t_j| \approx \Delta t$,
and $m$ are the magnitudes of any QSO in the database at those epochs. 
Hook \etal\ (1994) and Cristiani
\etal\ (1996) showed that the ensemble SF rises steadily for 
$\Delta t \lsim 2-3$~yr
and flattens at larger lags. Therefore we can regard the pulse life-time
to be  $1.5 \lsim \tau \lsim 3$~yr.

  The energy of the pulses can be estimated from fits of the form
$\sigma \propto L^{a} (1+z)^b$ to the data 
(CAT96). Note that the redshift term 
is analogous
to the wavelength term discussed in section 2.1. The 
proportionality constant of the $\sigma - L - z$ relationship 
is directly linked to the 
background fraction and to the luminosity of each pulse. Since
the energy and the luminosity of the pulses are related by the pulse 
life-time, the energy $\epsilon$ 
is found to be 
$1.5 \times 10^{50} \lsim \epsilon \lsim 6.3 \times 10^{51}$~erg,
within the limits of $f_{bck}$ and $\tau$ given in this section.

  From the total variable luminosity
of the QSOs $(1-f_{bck})\overline{L}$ and the energy of the pulses, we derive
that the rate of events is
$3 \lsim \nu \lsim 150$~yr$^{-1}$ for a $M_B=-25$~mag QSO,
within the limits of $f_{bck}$, $\tau$ and $\epsilon$ given above.

  Although the range of parameters is not tightly constrained,
it gives us an idea of the parameter-space to look for event
candidates.

\section{Search of 'pulses' in Seyfert nuclei}

Any search of isolated pulses of variation will be more satisfactorily carried
out in Seyfert~1 nuclei than in QSOs. Seyfert nuclei share all the properties
of QSOs across the electromagnetic spectrum, and are generally regarded as
lower luminosity counterparts of QSO activity in the nearby Universe. It is 
their lower luminosity ($M_V > -23$ mag) which makes them attractive for 
attempts to isolate pulses of variation, since the rate of events in these 
objects is predicted to be so much lower.

  In their study of the historical optical light curve of NGC~4151 
($\overline{M_B} \sim -20$~mag) Aretxaga \& Terlevich (1994a, hereafter AT94) 
draw the attention to a 
sharp pulse which raised from a deep minimum in 1970 and lasted for a few 
weeks, followed by a 
second peak of slow decline (a few years) with at least two secondary 
maxima --- see fig.1. The second peak comprises a $B$-band 
energy of 
$\epsilon \approx 3 \times 10^{50}$~erg.
They found some other similar events in the light curve of NGC~4151, 
reaching the 
conclusion that the pattern of the light curve could be originated by
the superposition  
of a basic unit of variation, isolated in the 1970--1973 event.

\begin{figure}

    \cidfig{4.0in}{87}{112}{488}{405}{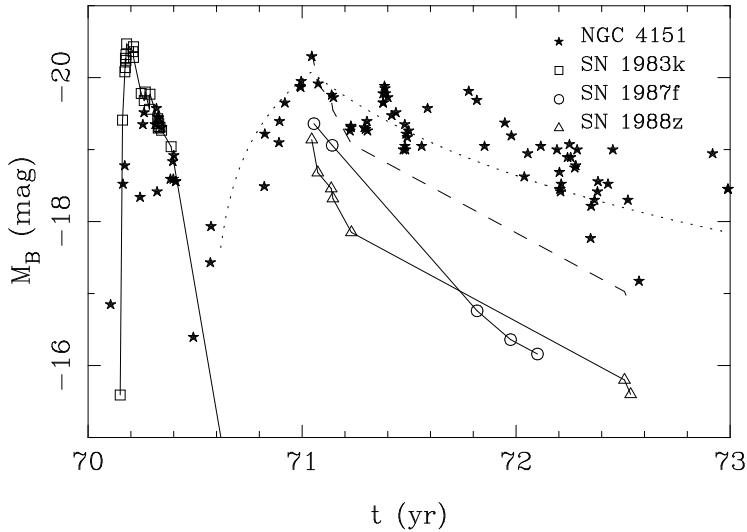}
    \caption{
Comparison of the net variation of the $B$-band light curves of
NGC~4151  and SN~1983K, SN~1987F and SN~1988Z.
The dotted line shows the theoretical $B$-band light curve of a cSNR
($\epsilon_{51}=3$, $n_7=1$). The
dashed line represents the light curve of SN~1988Z,  scaled to match the
second peak of the light curve of NGC~4151.
}
\end{figure}

  Perhaps the most impressive repetition of this kind of double-peak
event has been found in NGC~5548 ($\overline{M_B} \sim -21.1$~mag), 
tightly monitored by the AGN Watch
consortium since late 1988 (Korista et al. 1995 and references therein).
After a deep minimum in 1992 when the nucleus lost almost all its broad \Hb\
line (Iijima \etal\ 1992), the light curve developed a double-peak
event as that of NGC~4151, with secondary maxima in the second slow-decay
peak at similar time-scales as those of NGC~4151, and similar energies 
(Cid-Fernandes, Terlevich \& Aretxaga 1996). 
As in NGC~4151, previous events in the light curve where identified 
to approximately match the pattern of the double-peak event.

  NGC~4151 and NGC~5548 are by far the historically best followed up
Seyfert nuclei. But even in the much scarcely monitored low-luminosity 
($\overline{M_B} \gsim -18$~mag)
NGC~1516  the double-peak pattern 
can be detected  (see Alloin et al. 1986) where, however,
the time-scale of the slow-decay peak seems to be larger.

\section{The Starburst model: a physical Poissonian model}

The conventional wisdom is that AGN variability is produced by 
instabilities
in the accretion disk around a supermasive black hole (e.g. Rees 1984,
Wallinder et al. 1992). However, to date
there are no specific predictions in 
this model for the behaviour of the 
optical continuum. An important question in this respect is whether such 
instabilities should produce well defined patterns in the light curves,
or whether the number of events should be correlated with the intrinsic 
luminosity of the object, as discussed in the previous sections.

  The starburst model, reviewed by Cid Fernandes in this volume,
gives a natural explanation for those effects. 
In this context, the optical
variability observed in AGN is produced by
supernovae (SNe) which generate rapidly evolving compact
supernova remnants (cSNRs) due to the interaction of their ejecta with
the high density circumstellar environment created by their progenitor
stars.  During the SN~II phase, when the stellar cluster is 10--60~Myr old,
the bolometric luminosity is dominated by stars, while
the basic broad line region properties can be ascribed to the
evolution of cSNRs in a medium with densities $n \gsim 10^7$~\uniden\
(Terlevich \etal\ 1992). 
The observational evidence that most strongly supports this picture
is the striking similarities of the optical spectra and light evolution 
of AGN and cSNRs, 
like SN~1987F or SN~1988Z, the so-called 'Seyfert~1 
impostors' (Filippenko 1989).

\subsection{The basic unit of variation}

The two main features of the light
curve in fig.1, the rapid first peak (70.2) 
and the second slower-decay peak (71) with secondary maxima,
were compared with published light curves of SNe (AT94),
finding that the 70.2 peak was reproduced in luminosity and shape by
classical type~II SNe, like SN~1983K. The 71 peak had the same 
decline rate as cSNRs like SN~1987F and SN1988Z.
Although these objects seem to be less luminous than NGC~4151, 
they are believed to have a high intrinsic extinction 
since they are embedded in H~II regions. 
While there is no observational confirmation of cSNRs
developing secondary maxima in their light curves (the wiggles in NGC~4151
after 71), 
hydro-dynamical models of cSNRs show that the bolometric 
luminosity undergoes secondary peaks of about $10^{49}$--$10^{50}$~erg
associated with thin shell formation and shell-shell collision,
as well as time-unresolved rapid variability associated with cooling 
instabilities (Plewa 1995).

Therefore, {\it the basic unit of variation represented in fig.1 is associated
with the explosion of a SN (70.2 peak) which develops a cSNR (71 peak).}
A simple semi-analytical approximation of the light curve of a cSNR is given by 
$ L_B =    6 \times 10^{9} \LBsun \:  \e51^{7/8} \n7^{3/4} \:  
\left(t/t_{\mbox{\tiny sg }}  \right)^{-11/7}
$
\ \ for $t>t_{\mbox{\scriptsize sg}}$,
where
$
  \tsg = 0.62 \mbox{\ yr \ } \e51^{1/8} \n7^{-3/4} \mbox{\ \ }
$
is the cooling time, \e51\ is the bolometric 
energy released in each remnant in units of
$10^{51}$~erg, 
and \n7\ is the circumstellar density in which the cSNR
evolves, in units of $10^7$~\uniden (Terlevich et al. 1992). 
This light curve  for $\e51=3$ and $\n7=1$ is represented 
in fig.~1 with a dotted line.

\subsection{Intrinsic scaling relationships}

  The $B$-band luminosity arising from a coeval stellar cluster at its SN~II
explosion phase is mainly due to the contribution of
main sequence stars and SNe. 
The SN
rate (\SNrate) and the optical luminosity coming from stars (\LBstar) are
related along the lifetime of this phase by
$
  \SNrate / \LBstar \approx 2 \times 10^{-11}
  \mbox{ \ \ yr$^{-1}$ \LBsun$^{-1}$\ ,}
$
almost independently of the IMF and age
of the cluster (AT94). 
From this, we deduce that the mean total luminosity 
of the cluster is related to the SN rate by
\begin{equation}
  \overline{\LBclu} 
  \sim 5 \times 10^{10} \: \frac{\SNrate}{\mbox{yr$^{-1}$}} \: 
(\epsilon_B + 1 )
  \mbox{ \ \LBsun \  \ \ ,}
  \label{eq:lum}
\end{equation}
where \eB\ is the mean B-band energy released in each SN remnant, in
units of $10^{51}$~erg.
An estimation of \eB\ can be obtained from the observed time-averaged
equivalent width of \Hb,
\begin{equation}
  \overline{\ewHb} \sim 320 \mbox{\AA\ \ } \frac{\eB}{1+0.17\eB}
  \mbox{\ \ ,}
 \label{eq:ewHb}
\end{equation}
which is also independent of the age, mass or IMF of
the cluster, but is weakly dependent on the adopted 
bolometric correction of cSNRs (AT94). For SN~1987F and SN~1988Z we find
$\eB/\e51 \approx 0.12$.
The near-constancy of the time-averaged equivalent with of \Hb\ is a central 
result of the starburst model. If the energy per supernova has a universal 
value, then the equivalent width of \Hb\ in AGN with broad lines should be in 
a narrow range of values.

\subsection{Parameters of the model}

From these expressions we can deduce that {\it the 
four parameters that characterize a Poissonian process
can be described with just one functional parameter in the starburst model}
(Aretxaga, Cid Fernandes \& Terlevich 1996, ACT96 hereafter):
\\ \hspace*{1cm} 1. \ 
The main source of non-variable background luminosity comes from the 
stars in the cluster. The stellar luminosity is directly linked to the
rate of events by a universal value. This makes for approximately half
of the total luminosity of the nucleus. 
\\ \hspace*{1cm} 2. \ 
The energy of the events 
is $ \epsilon_B \sim 5 \times 10^{50}$~erg, in order to satisfy
eq.2 with the observed values of the equivalent width 
of \Hb\ in QSOs, $\overline{\ewHb} \approx 100$~\AA\ (Osterbrock 1991).
Values of the kinetic energy released
in a SN explosion of up to $3 \times 10^{51}$~erg are measured 
in type II SNe (Branch \etal\ 1981).
\\ \hspace*{1cm} 3. \ 
The rate of events is linked to the total luminosity of the objects
by eq.1, and also depends on the energy of the events.
For a $M_B=-25$~mag source, the deduced rate is $\approx 20$ SN/yr.
\\ \hspace*{1cm} 4. \ 
The shape of the events is given by the SN$+$cSNR light curve
represented in fig.1, which depends on the energy ($\eB \approx 0.5$)
and the characteristic time-scale of the pulses (\tsg).
This last parameter actually controls the shape of the light
curve, and remains free.  However, its value can be constrained to a narrow
band. The values of \tsg\
        found to reproduce well isolated peaks in the light curves of NGC~4151
and NGC~5548
        are \hbox{260--280 days} (AT93, AT94), 
	but since high luminosity QSOs may have higher
        metallicities (Hamann \& Ferland 1993), the evolution of
        their cSNRs could be substantially faster, as cooling 
        rates increase with metallicity. 

\newpage
\subsection{Comparison with the data}

It should be noted the good agreement between 
the predicted values derived from
the starburst model and the empirically determined values
in section~2.2 : $f_{bck} < 0.7$,
$1.5 \times 10^{50} \lsim \epsilon \lsim 6.3 \times 10^{51}$~erg,
and
$1.5 \lsim \tau \lsim 3$~yr 
where the life-time of the pulses and the cooling-time are related by
$\tau \approx 5 \tsg$.

  Most importantly, the luminosity--SN rate relationship of eq.1
gives the zero-point of the variability-luminosity relationship. Monte Carlo
simulations of stellar clusters undergoing SN explosions  show
that this scaling matches the zero-point of the 
variability-luminosity relationship of QSOs once single pass-band data 
are corrected for 
wavelength-effects (ACT96). The agreement is good for models with time-scales
$85 \lsim \tsg  \lsim 280$~days and objects spanning 7 magnitudes.
The analysis of ensemble light curves doesn't show any clear variations
of \tsg\ with redshift, although some evolution is expected and this 
might be reflected by
the goodness of models with different \tsg\ values.

  The asymmetry of the cSNR light curves and the apparent time-reversibility
of QSO light curves was pointed out as a potential problem by M. Malkan 
during this conference. 
To address this issue, in collaboration with S. 
Cristiani we run some simple tests on the $B$-band light curves of
180 QSOs monitored for 18 epochs over 10 years 
(Cristiani et al. 1996). The ensemble skewness of the variations of
the sample was measured,
finding that within the error bars
the number of increases is the same as  the number of decreases
over the different luminosity variation intervals. 
The same is true
for theoretical light-curves generated using the simple semi-analytical
solution of fig.1 (dotted line) once sampling and observational errors
are introduced. One should note that even error-free infinitely-sampled 
light curves 
built up from the superposition of cSNR units look very symmetric
due to the 
high number of events superposed --- see fig.2 for a $M_B=-25$~mag 
theoretical light curve --- much more when sampling (just 20 epochs
scattered in 10 years) and observational errors (about 0.1~mag for 
most samples) are introduced. 

\begin{figure}
    \cidfig{5.5in}{1}{178}{570}{333}{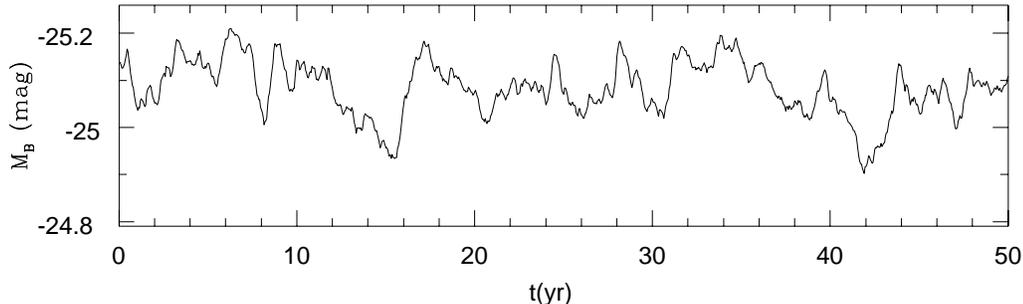}
    \caption{
$B$-band light curve of a cluster undergoing SN explosions
at a rate $\SNrate = 20$~\uniSNrate. The time-scale of the cSNRs
is $\tsg=280$~days, which gives the highest possible asymmetry.
}
\end{figure}

  For typical QSOs with luminosities between $-24$ and $-27$~mag, 
the SN rate varies between 5 and 100~\uniSNrate\ and
the masses of the postulated coeval stellar clusters vary from  
$2\times10^{10}$ to  $3\times 10^{11}$~\Msun\ for a solar neighbourhood 
IMF (AT94).  The mass estimation changes with the
IMF and the age of the cluster. The derived masses
are well inside the hypothesis of Terlevich \& Boyle (1993) that the QSO 
phenomenon might correspond to the formation of the cores of nowadays normal
elliptical galaxies  at $z\gsim 2$.. 

  For the Seyfert~1 nuclei NGC~4151 and NGC~5548 the 
SN rates derived from the luminosity 
are 0.2 and 0.3~\uniSNrate\
respectively, the energy of the events derived from \ewHb\ 
is $\eB \approx 0.5$ and the
time-scale of the pulses can be deduced from isolated pulses of variation
like the one in fig.1 to be $\tsg \approx 280$~days. Monte Carlo simulations
of models with these parameters reproduce in detail the main weak-scale
characteristics of the light curves, their rms and power spectra (AT93, AT94).

  The low rate of events also explains the 
temporal disappearance or appearance of broad lines in objects which 
are traditionally classified as Seyfert~1 or Seyfert~2 and LINERs, 
respectively. Quiescent stages with undetectable broad lines occur at deep
photometric minima. The transitions could be produced in periods of time 
in which the existing remnants are too old to produce broad lines
and the probability of a new explosion which refuels the broad line 
region is very low. The continuum in these epochs is still blue because 
the light is dominated by young main sequence stars.
A comparison between the probability for this process to happen and 
recorded occurrences of type transitions shows that all the AGN
which have changed types of activity are within the threshold
$\overline{M_B} \gsim -22.5$~mag (Aretxaga \& Terlevich
1994b). For objects of higher luminosity the probability of a new 
explosion at any given time would be too high to allow the broad lines to 
disappear.

Although the optical light curves of radio-quiet AGN
follow the predictions of the starburst model, not all AGN follow that
pattern. For example, in the historical light curve of the 
OVV QSO 3C345 (Babadzhanyants et al. 1995) a very similar
'basic unit of variation' can also be
isolated, but the total luminosity of the source
$M_B \approx -26.6$~mag predicts a too high rate of events to account for
the 3~mag amplitude variations in $B$-band. Blazars and OVV QSOs 
are a minor brand of radio-loud AGN in which other mechanisms should be 
in play.

\section{Conclusions}

The optical variability of radio-quiet AGN is 
found to be consistent with simple Poissonian processes.
A potential basic unit 
of variation is identified in Seyfert nuclei. The energies and time-scales
of this unit are similar to those of nearby SNe and cSNRs.
The energies, time-scales and rates of the events empirically found in
large databases of QSOs are consistent with those expected from the starburst
model of AGN. Two strong predictions are confirmed by these objects: 
(a) the number of events scales with luminosity, with the zero-point 
derived from stellar evolution (eq.1)
and (b) if the energy per supernova has a universal 
value, then the equivalent width of \Hb\ in AGN with broad lines should be 
constrained to a narrow range of values (eq.2).

\vspace*{0.5cm}
\ni
ACKNOWLEDGMENTS\\
It is a pleasure to acknowledge 
R. Cid Fernandes, S. Cristiani,
L. Sodr\'e and R. Terlevich with whom most of the work presented here has 
been done, and I. Salamanca for comments on an early draft of this paper. 
IA's work is supported by the EEC fellowship ERBCHBICT941023.

\end{document}